# A vectorial Doppler effect with spatially variant polarized light fields

Liang Fang[1+], Zhenyu Wan[1+], Andrew Forbes[2], Jian Wang[1*]

[1]Wuhan National Laboratory for Optoelectronics, School of Optical and Electronic Information, Huazhong University of Science and Technology, Wuhan 430074, Hubei, China.

[2]School of Physics, University of the Witwatersrand, Private Bag 3, Johannesburg 2050, South Africa

[+] These author contributed equally to this work.

[*] Correspondence to: jwang@hust.edu.cn

**The Doppler effect of light was implemented by interference with a reference wave to infer linear velocities in early manifestations, and more recently lateral and angular velocities with scalar phase structured light. A consequence of the scalar wave approach is that it is impossible to deduce the motion direction and position of moving targets directly. Here we overcome this limitation with vectorially structured light, exhibiting spatially variant polarization, allowing both the position and velocity of a particle to be robustly determined by a new vectorial Doppler effect. We use this to successfully implement real-time monitoring of instantaneous angular position and velocity of a complicated moving particle, and interpret its physical mechanism as evidence of time-reversal symmetry breaking in classically entangled light fields. Our demonstration can easily be extended with arbitrary vector structured fields for complete motion tracking of non-trivial trajectories, offering new vectorial Doppler metrology for universal motion vectors.**

The Doppler effect is a universal wave phenomenon. Since its first discovery in 1842, it has been widely applied to acoustic and optical metrology in various subjects, such as astronomy, oceanography, medicine, engineering, etc [1-18]. In particular, as for light wave, because of its ultra-high velocity, large bandwidth, and perfect directionality, the Doppler effect of light has spurred a myriad of applications, from cooling atoms to monitoring traffic flow. This effect originates from the relative motion between a wave source and an observer, resulting in a frequency shift with respect to wave frequency. In early manifestations, the common linear Doppler effect of wave was used for deduction of linear velocities along the direction of wave. More recently, the rotational Doppler effect of helical phase wave was revealed to allow determination of rotational motion for both acoustic and light waves[3-8].

Throughout the development history of the Doppler effect, as well as its wide applications, it has always been based on scalar waves with up or down frequency shift. The resulting Doppler shift is easily sensible and detectable for scalar waves with low frequency, such as water and acoustic waves[8]. However, as for the light (electromagnetic) wave, its Doppler shift relative to the ultra-high inherent frequency (up to several hundreds of THz) cannot be detected by common intensity-sensitive detectors directly. Thus, extracting the Doppler shift of light is usually implemented by interference with a coherent reference light. However, this implementation only allows the scalar speed determination for the movement measurement, but cannot get the direction information, unless the additional means of dual-frequency or heterodyne detection is adopt[19-21].

Apart from the well-established scalar (amplitude and phase) degrees of freedom, light also has a state of polarization (SoP) that describes the oscillated electric field in the plane perpendicular to the propagation direction. In particular, for structured light with tailored spatial polarization, instead of the uniform SoP of scalar optical fields, this new family of vectorial polarization fields (VPFs) is characterized by spatially variant polarized light fields across the transverse plane, and may form polarization vortices (singularities)[22-26]. The most typical VPFs are azimuthal and radial polarization fields[19], and even higher-order VPFs, where the SoP can be well illustrated by the equator on the higher-order Poincaré sphere[27,28]. These cylindrical VPFs previously attracted a lot of attention as eigenmodes in optical fibers[29]. In the past decade, from fibers to free space and even to integrated devices, such fields have gained increasing interest and given rapid development in a diversity of applications in laser material processing[30], optical tracking[31,32], microscopy[33], particle acceleration[34], classical entanglement and sensing[35-38], etc.

Here we reveal a vectorial Doppler effect that exploits such vectorially structured light fields. We exploit the polarization degree of freedom to extract directional information by following the Doppler response of the particle as it traversing a spatially structured polarization field. We outline the concept theoretically with the aid of well-known cylindrical VPFs, and use this class of structured light to demonstrate proof-of-principle experiments. Using arbitrary angular trajectories, we show that we can robustly reconstruct the instantaneous velocity vector in magnitude and direction of a moving particle, as well as its instantaneous position, which may pave the way for real-time monitoring and tracking of in principle any trajectory in 3D.

**Results**

**Concept and principle.** The concept and principle of our vectorial Doppler effect with spatially variant polarized light fields were illustrated in Fig. 1, where a representative $HE_{31}$-like VPF (Figs. 1a and 1c) was used for conceptual detection of the angular velocity of a moving particle. Generally, the particle without birefringence does not change the SoP when scattering/reflecting light based on the Mie's theory[39]. Thus, when moving within a spatially variant polarized light field, the particle will reflect/scatter the local light with the same polarization. The scattered/reflected polarized light by the moving particle from this field can be interpreted as Doppler polarization signal (DPS). If the polarization distribution of this field is spatially periodic (for example, shown as Figs. 1a and 1c), the DPS can be written as a Jones vector,

$$\vec{E}(v,t) \simeq A \cdot \begin{bmatrix} \cos(k\gamma vt - \alpha) \\ -\sigma \cdot \sin(k\gamma vt - \alpha) \end{bmatrix}, \qquad (1)$$

where $A$ is the real amplitude of the polarized light. Here we assume that the particle moves within VPFs in a uniform intensity region, thus giving the constant $A$. $k = 2\pi/\lambda$ is the wavenumber of light with $\lambda$ being the wavelength. $v$ denotes the linear velocity of the moving particle. $\gamma$ stands for the small angle between the Poynting vector of incident light and the observation direction, which determines the spatial period of polarization orientation ($\Lambda = \pi/k\gamma = \lambda/2\gamma$) of the VPFs. $\alpha$ is related to the initial polarization orientation (see supplementary materials). Note that the sign $\sigma$ takes '+1' or '-1', describing two opposite chirality of DPS, derived from the polarization orientation of fiber eigenmodes $HE_{n,1}$ and $EH_{n,1}$ like modes, respectively, where n denotes the fold number of rotational symmetry.

If the moving particle reverses its velocity vector ($v \rightarrow -v$), the resulting DPS becomes

$$\vec{E}(-v,t) \simeq A \cdot \begin{bmatrix} \cos(k\gamma vt + \alpha) \\ \sigma \cdot \sin(k\gamma vt + \alpha) \end{bmatrix}. \qquad (2)$$

From Eqs. (1) and (2), one can see that no matter what values $\alpha$ takes, the DPSs always keep $\vec{E}(-v,t) \neq \vec{E}(v,t)$. The DPSs show remarkable chirality inversion when reversing the motion direction of the particle (see Figs. 1b and 1d). These two-dimensional DPSs belong to the distinct property of vectorial Doppler effect with spatially variant polarized light fields that enables the distinguishability of motion direction of a moving particle. By contrast, the classical Doppler effect based on scalar optical fields shows one-dimensional Doppler intensity signals, incapable of distinguishing the motion direction (see supplementary materials).

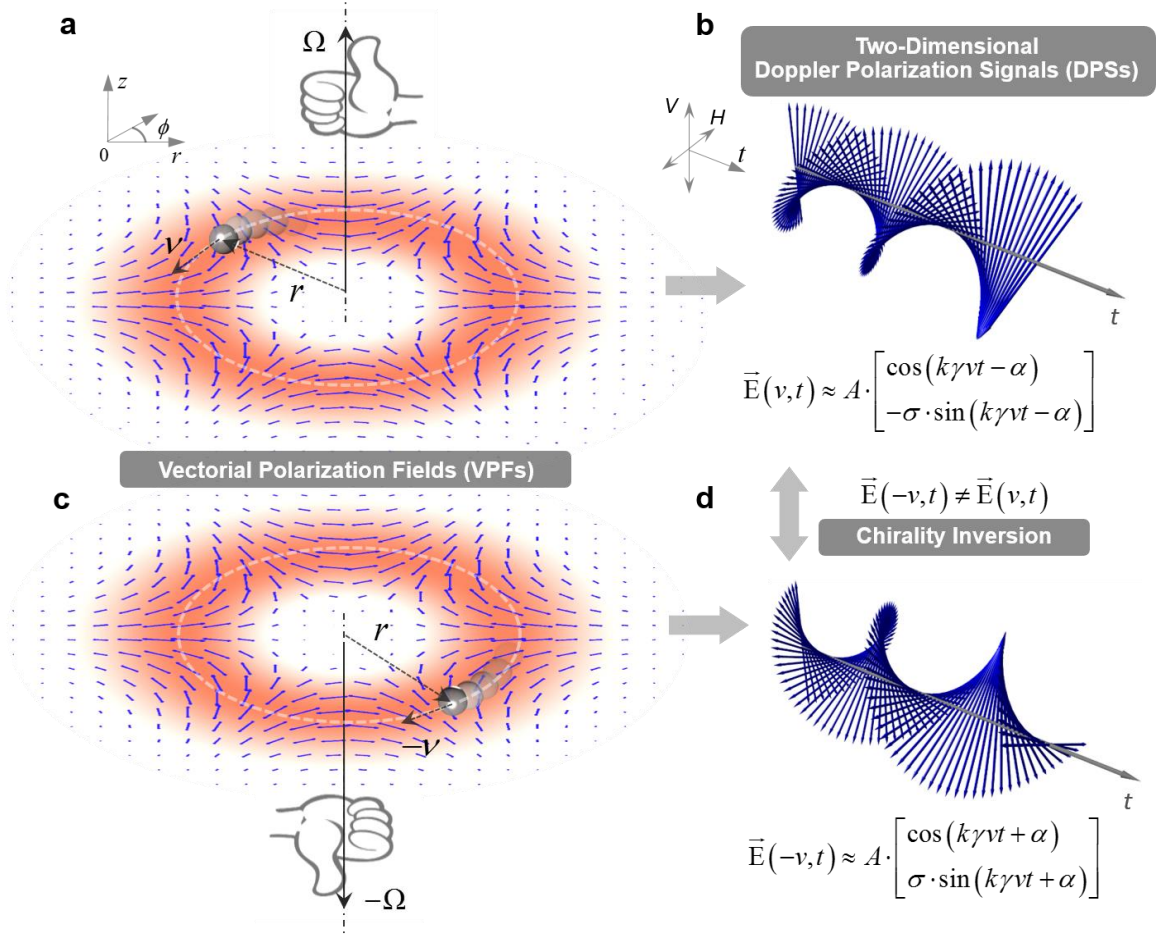

**Fig. 1** Vectorial Doppler effect with spatially variant polarized light fields used for robust determination of motion vector. **a**, **c** The cylindrical $HE_{31}$-like field as a representative VPF is conceptually shown for direct detection of a moving particle with plus and minus angular velocity vectors, respectively. **b**, **d** The two-dimensional Doppler polarization signals (DPSs) scattered from the VPFs show chirality inversion for plus and minus angular velocity vectors. The two-dimensional DPSs carry full information (magnitude, direction) of velocity vectors, which is not available by classical Doppler effects based on scalar optical fields. The blue arrows denote state of polarization. $r$: radius; $v$: linear velocity; $\Omega$: angular velocity.

In practice, one can exploit the chirality of DPS through two polarizers to distinguish the motion direction (or the sign of velocity vector). The phase relation of Doppler signals after passing through the polarizer with the polarizing angle $\theta_j$ (with respect to the $x$ axis) can be deduced as

$$I_j(v,t) = \left| \begin{pmatrix} 1 & 0 \\ 0 & 0 \end{pmatrix} \cdot \begin{pmatrix} \cos\theta_j & \sin\theta_j \\ -\sin\theta_j & \cos\theta_j \end{pmatrix} \cdot \vec{E}(v,t) \right|^2 \propto 1 + \cos\left[2(k\gamma vt + \sigma\theta_j - \alpha)\right], \quad (3)$$

where $j = 1, 2$ indicates polarizers 1 and 2 ($P_1$, $P_2$). It shows that the initial phase of this Doppler signal is dependent of the polarizing angle $\theta_j$ of the polarizer. Under a reversed movement ($v \to -v$), this Doppler signal can be rewritten as $I_j(-v,t) \propto 1 + \cos\left[2(k\gamma vt - \sigma\theta_j + \alpha)\right]$. As for single polarizer, obviously, the detected Doppler signals satisfy $I_j(v,t) = I_j(-v,t)$ under the random phase of $\alpha = \sigma\theta$, thus resulting in incapable of distinguishing the motion direction. This case actually falls into the traditional Doppler velocimetry with Doppler intensity signal based on scalar optical fields (see supplementary materials).

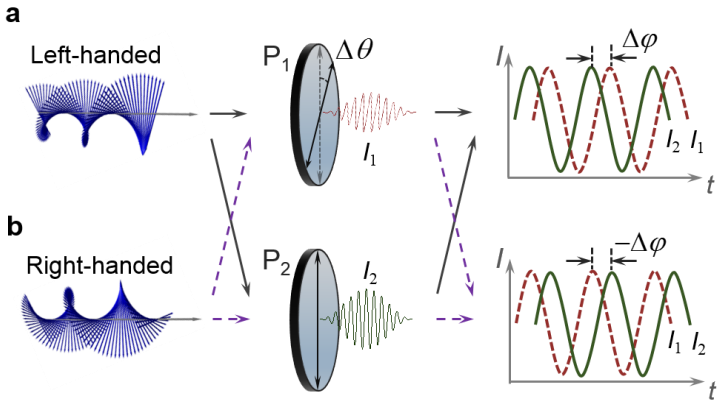

**Fig. 2** Two polarizers used for detection of two-dimensional DPSs to determine both the magnitude and direction of motion vector. **a** The left-handed DPS passing through two polarizers ($P_1$ and $P_2$) gives a fixed Doppler phase shift between Doppler intensity signals 1 and 2 ($I_1$ and $I_2$). **b** The right-handed DPS passing through $P_1$ and $P_2$ gives an opposite Doppler phase shift between $I_1$ and $I_2$, relative to the case of left-handed DPS.

If two polarizers ($P_1$, $P_2$) are employed to detect the DPS synchronously (Fig. 2), a relative phase difference (RPD) forms between the gotten Doppler intensity signals $I_1$ and $I_2$. From Eq. (3), the

RPD between them can be written as $\Delta\varphi = \varphi_2 - \varphi_1 = \text{sign}(v) \cdot 2\sigma\Delta\theta$, where sign() stands for the sign function, and $\Delta\theta = \theta_2 - \theta_1$. The factor of 2 in RPD derives from the 2-to-1 homeomorphism between the physical SU(2) space of the light beam and the topological SO(3) space of the Poincaré sphere[27]. It clearly shows that the RPD is dependent of the charity of DPSs (Figs. 2a and 2b). In return, the direction of the velocity vector can be distinguished through this RPD that can be defined as new Doppler phase shift. The magnitude of the velocity vector can be gotten through the common Doppler frequency shift $|\Delta f| = |k\gamma v/\pi|$ as experimental observable. This inherently direction-sensitive vectorial Doppler effect benefits from the two-dimensional DPS of spatially variant polarized light fields. Furthermore, based on the principle of vectorial Doppler effect, significantly, one can get a real-time monitoring for instantaneous position and velocity of a moving particle with variable motion (see Methods).

**Experiment proof.** We first demonstrate the robust detection of angular velocity vector of a rotating particle by our vectorial Doppler effect with a cylindrical VPF (Fig. 3a). In principle, for a particle that rotates around an axis with a linear velocity $\vec{v}$, perpendicular to the radial vector $\vec{r}$, the angular velocity vector (pseudovector) can be defined by $\vec{\Omega} = (\vec{r} \times \vec{v})/r^2$. The magnitude of the angular velocity is determined by the rate of change of the angular position of the rotating particle, usually expressed in radians per second, i.e. $\Omega = d\phi/dt$. The direction (plus or minus) of the angular velocity is either up or down along the axis of rotation by the right-hand rule (Figs. 1a and 1c). In general, the cylindrical VPF can be generated by many approaches, such as the superposition of two light components with different twisted phase structures $\pm\ell$ and circular polarization $\sigma = \pm 1$ [27,28], where $\ell$ represents the topological charge number. In this case, the angle $\gamma = \ell/kr$ in Eqs. (1)-(3) represents the skew angle of the Poynting vector of the twisted light around the propagation axis[40,41], and the linear velocity is $v = r\Omega$. In this case, the DPS in Eq (1) can be written as $\vec{E}(v,t) \simeq A \cdot [\cos(\ell\Omega t - \alpha), \ -\sigma \cdot \sin(\ell\Omega t - \alpha)]^T$, where the superscript 'T' denotes matrix transposition.

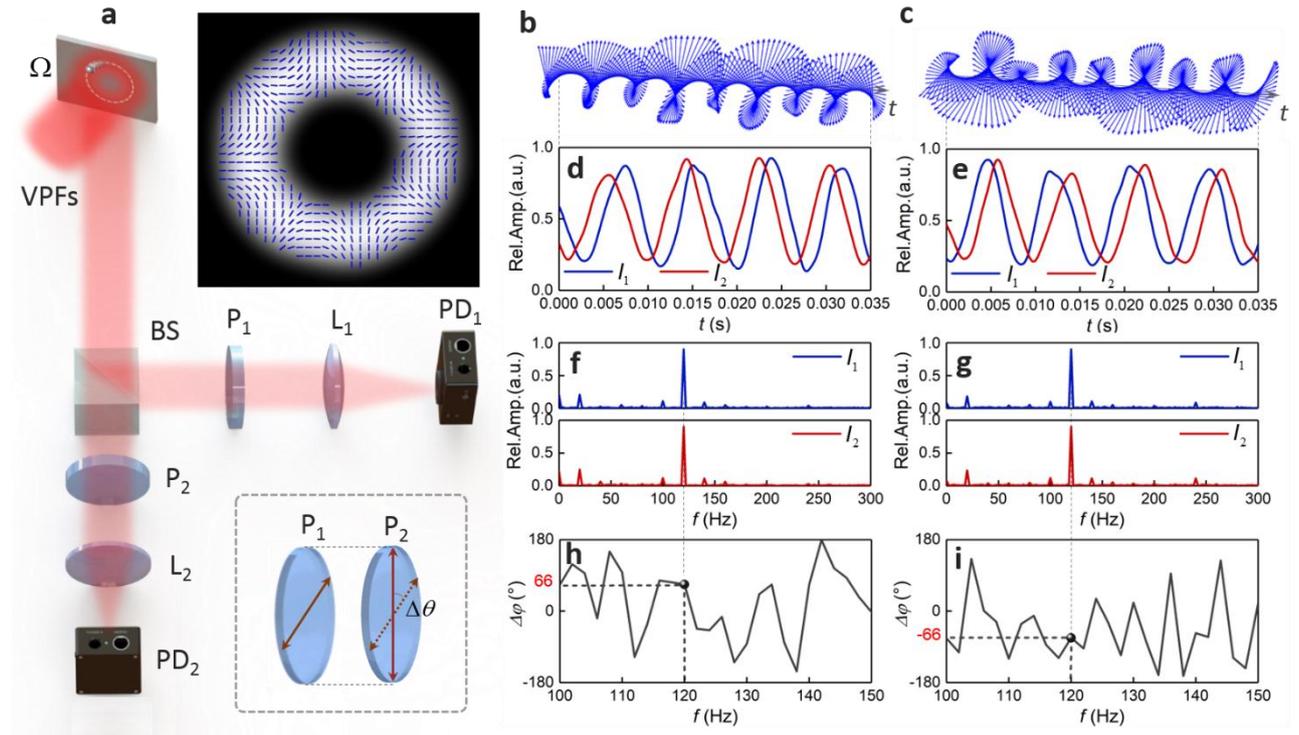

**Fig. 3** Experimental configuration and measured results of vectorial Doppler effect. **a**, The cylindrical HE$_{41}$-like VPF illuminates a rotating particle, and the reflected DPS is split into two paths by a beam splitter (BS), then filtered by two polarizers (P$_1$, P$_2$) with a polarizing angle difference of $\Delta\theta$ (down inset), and finally detected by two photodetectors (PD$_1$, PD$_2$). The up inset shows the measured intensity and polarization distributions of the high-quality HE$_{41}$-like VPF generated in the experiment. L$_1$, L$_2$: lens. **b**, **c** Recovered DPSs show right-handed (as defined from the point of view of the origin) and left-handed under two opposite angular velocities. **d**, **e** Measured Doppler intensity signals by PD$_1$ and PD$_2$, respectively. **f-i** Doppler frequency spectra acquired by fast Fourier transform (FFT) for the recorded Doppler intensity signals. **f**, **g** Amplitude-frequency spectra. **h**, **i** Phase-frequency spectra as RPDs between the recorded Doppler intensity signals 1 and 2. **b**, **d**, **f**, **h** $\vec{\Omega} = 40\pi$ rad/s. **c**, **e**, **g**, **i** $\vec{\Omega} = -40\pi$ rad/s.

In the experiment, we generated a HE$_{41}$-like cylindrical VPF to detect the angular velocity vector of a rotating particle (Fig. 3a) that was mimicked by a digital micromirror device (DMD) (see Methods and supplementary materials). The cylindrical VPF pattern needs to be adjusted to match the rotational trajectory of the particle when illuminating this rotating particle. The reflected DPSs (Figs. 3b and 3c) by this rotating particle with two opposite angular velocities were split into two paths, and then filtered by polarizers 1 and 2 (P$_1$, P$_2$) in each path with the polarizing angle difference about $\Delta\theta \approx 33°$. After this, the corresponding Doppler intensity signals 1 and 2 were collected by the photodetectors 1 and 2 (PD$_1$, PD$_2$), as plotted in Figs. 3d and 3e, respectively. The Doppler frequency spectra in Figs. 3f-3i were acquired by fast Fourier transform (FFT) for the Doppler intensity signals, including both amplitude-frequency spectra (Figs. 3f and 3g), as well as phase-frequency spectra (Figs. 3h and 3i). For convenience, we presented the phase-frequency spectra (Figs. 3h and 3i) plotted as the RPDs between the Doppler intensity signals 1 and 2. From the experimental results, one can clearly see that the DPSs reverse their chirality (Figs. 3b and 3c) and thus give rise to the delay/advance in time between Doppler intensity signals 1 and 2 (Figs. 3d and 3e) under opposite angular velocities. All the measured amplitude-frequency spectra show a fixed Doppler frequency shift (Figs. 3f and 3g). However, the corresponding Doppler phase shifts give opposite values (Figs. 3h and 3i), i.e. $\Delta\varphi \approx 66°$ for $\vec{\Omega} = 40\pi$ rad/s, while $\Delta\varphi \approx -66°$ for $\vec{\Omega} = -40\pi$ rad/s, in accordance with the theories.

With the same twisted phase structure basis forming the HE$_{41}$-like VPF, the cylindrical EH$_{21}$-like VPF was also employed to implement the same experiment. The comparison results show that the RPDs at the Doppler shift peaks reversed the symbols compared to the results using HE$_{41}$-like VPF (see supplementary materials). We further demonstrated the measured results using cylindrical VPFs with different orders (TM$_{01}$, HE$_{21}$, EH$_{21}$, HE$_{41}$, EH$_{41}$, HE$_{61}$, EH$_{61}$, HE$_{81}$, EH$_{81}$, HE$_{10,1}$), as plotted in Fig. 4a, as well as the measured results using EH$_{61}$-like VPF under different angular velocities (see supplementary materials). The experimental results verified that the corresponding mode-order ($\ell$) and forms ($\sigma$) of cylindrically VPFs play the same roles with the magnitude and direction of angular velocities to determine the corresponding Doppler frequency and phase shifts.

We also present the experimental results of the variation ($\Delta\varphi$) Doppler phase shifts with the polarizing angle difference ($\Delta\theta$) when detecting the DPSs using two polarizers (Fig. 4b). The Doppler phase shifts change twice as much as the polarizing angle difference and the sign of them also depends on the types of employed VPFs (HE or EH like VPFs), which is well consistent with theories ($\Delta\varphi = \text{sign}(\Omega) \cdot 2\sigma\Delta\theta$). Note that the polarizing angle difference should be avoided around $\pm 90°$ in the experiment, since the resulting Doppler phase shifts between $\pm 180°$ and $\mp 180°$ when detecting two opposite velocity vectors are equivalent in Doppler phase-frequency spectra so that cannot be effectively distinguished. All the measured results verifies that the spatially variant polarized light fields possess an inherently direction-sensitive (vectorial) Doppler effect that enables robust determination of velocity vectors of moving targets.

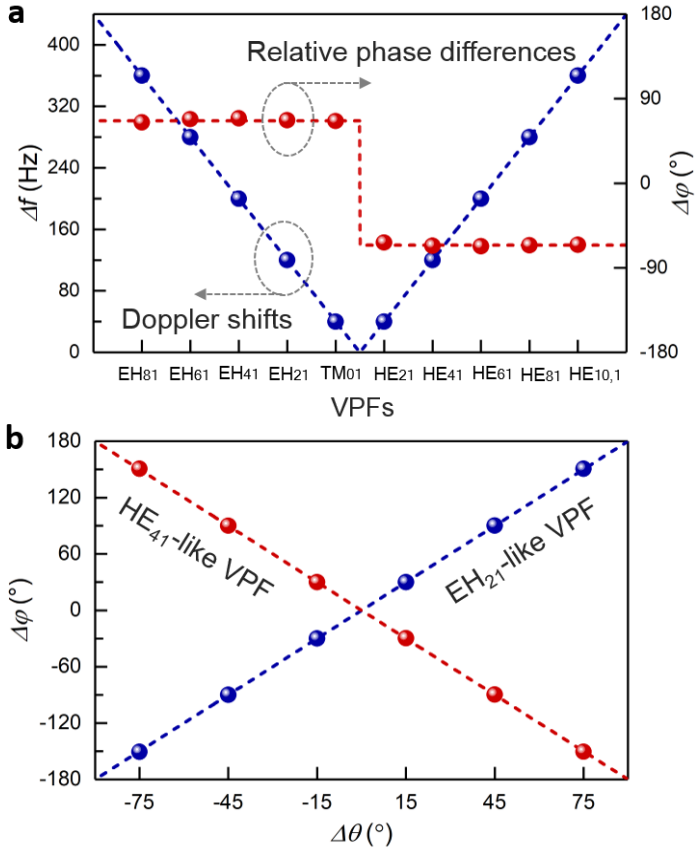

**Fig. 4** Measured results of vectorial Doppler effect using cylindrical VPFs with different orders. **a** Doppler frequency shifts and Doppler phase shifts versus different orders of cylindrical VPFs ($TM_{01}$, $HE_{21}$, $EH_{21}$, $HE_{41}$, $EH_{41}$, $HE_{61}$, $EH_{61}$, $HE_{81}$, $EH_{81}$, $HE_{10,1}$). **b** Doppler phase shifts versus polarizing angle differences between two polarizers used to detect the DPSs. The angular velocity is $\vec{\Omega} = -40\pi$ rad/s. The dots represent measured results and the dashed lines denote theories.

**Real-time monitoring.** We further demonstrate the capacity of real-time monitoring for instantaneous position and velocity of a moving particle with a complex motion vector based on our vectorial Doppler effect with spatially variant polarized light fields (see Methods). Here we first use a higher-order cylindrical VPF to implement the real-time monitoring of angular position and velocity of a representative pendulum motion. The particle also mimicked by DMD moving in such state rotates around the beam axis of a $HE_{19,1}$-like VPF (see Fig. 5a). Despite, strictly speaking, not being a harmonic motion for the pendulum motion with a pivot angle, the moving particle here was approximately controlled as the state of harmonic motion. The pendulum length was set as about 2 mm in line with the radius of cylindrical VPF. In this case, the moving particle with harmonic motion will reflect the local polarized light from the higher-order cylindrical VPF to a non-uniform DPS (Fig. 5b). This DPS carries the instantaneous motion information of the pendulum motion. The density of time-varying polarization variation of DPS shows the magnitude variation of velocity vector, while the chirality of DPS indicates the direction information, and especially, the chirality inversion implies the direction reverse of the movement.

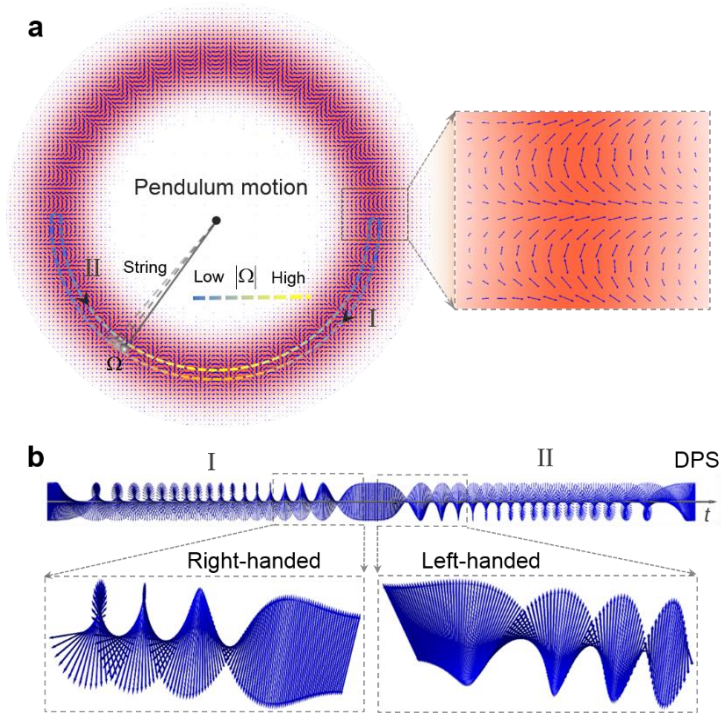

**Fig. 5** Real-time monitoring of a complicated moving particle based on vectorial Doppler effect with spatially variant polarized light fields. **a** A moving particle with pendulum motion is illuminated by the higher-order cylindrical $HE_{19,1}$-like VPF. The inset shows zoom-in details of spatial polarization distribution of the $HE_{19,1}$-like VPF. **b** DPS (Simulated) reflected from the higher-order cylindrical VPF can be available to retrieve the instantaneous position and velocity vector information of the moving particle.

In the experiment, after filtering the DPS (Fig. 5b) through two polarizers, we got the Doppler intensity signals (Fig. 6a). According to the principle of real-time monitoring (see Methods), we successfully obtained the instantaneous angular positions (Fig. 6b) and angular velocities (Fig. 6c) of the particle with pendulum motion. More generally, we also experimentally demonstrated the real-time monitoring of the moving particle with random rotation around the beam axis (Fig. 7). All the measured results verify the feasibility of real-time monitoring of a complicated moving particle based on vectorial Doppler effect with spatially variant polarized light fields.

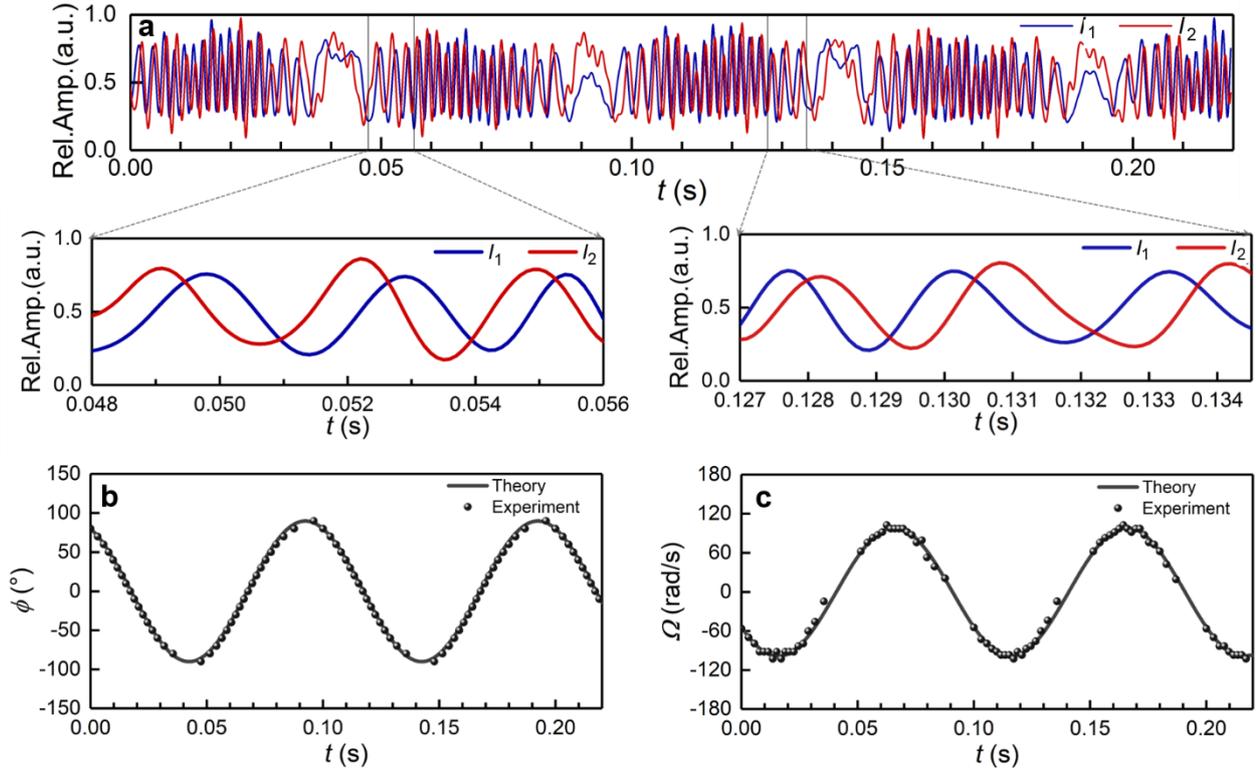

**Fig. 6** Measured results of real-time monitoring of a moving particle with pendulum motion using a HE$_{19,1}$-like VPF. **a** Measured Doppler intensity signals after filtering the DPS (Fig. 5b) by two polarizers. The insets are zoom-in details of the measured Doppler intensity signals in two different time windows. **b** Retrieved instantaneous angular positions. **c** Retrieved instantaneous angular velocities. The dots represent experiments and the solid lines denote theories.

In the real-time monitoring of a complicated moving particle, there will be inevitably few peaks in the Doppler intensity signals from the spatially variant polarized light fields when the velocity is close to zero, resulting in fewer measured data points and thus reducing the resulting resolution (see Figs. 6c and 7c). Even so, the temporal resolution of the real-time monitoring can be improved by decreasing the spatial period $\Lambda = \pi/k\gamma = \lambda/2\gamma$ of the spatially variant polarized light fields. In the practical monitoring, it means that one can generally enlarge the observation angle $\gamma$ or use the laser with shorter wavelength to improve the temporal resolution. Especially, as for the cylindrical VPFs ($\gamma = \ell/kr$), this measures correspond to increase the more order ($\ell$) or reduce the size ($r$) of cylindrical VPFs. Because when using this kind of vector structured light, the Doppler response of the moving particle interacting with the subtler spatial polarization period can give denser Doppler signals available to embody the subtler variation of the movement. Nonetheless, it is noteworthy that no matter how smaller periodic polarization distribution is used, the size of the targeted particle should be small enough relative to the period of spatial polarization, otherwise the DPS cannot be distinguished. Therefore, there is a trade off between the temporal resolution (VPF order) and the relative size of VPF for an effective real-time monitoring of a moving particle in practice.

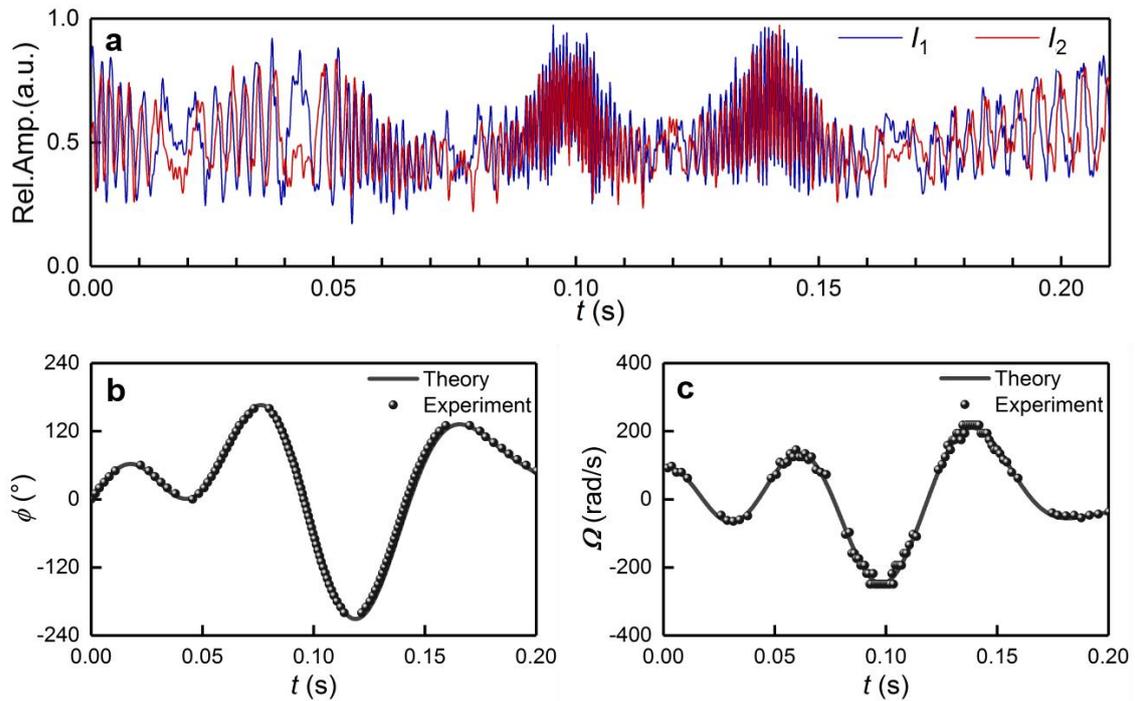

**Fig. 7** Measured results of real-time monitoring of a moving particle with random rotation using a $HE_{19,1}$-like VPF. **a** Measured Doppler intensity signals after filtering the DPS (Fig. 5b) by two polarizers. **b** Retrieved instantaneous angular positions. **c** Retrieved instantaneous angular velocities. The dots represent experiments and the solid lines denote theories.

## Discussion

Here we proposed a new vectorial Doppler effect with the spatially variant polarized light fields, characterized by new two-dimensional DPSs that can provide an additional RPD between different linear polarization components. Such RPD can be regarded as new Doppler phase shift, as an important supplementary of the conventional Doppler effect based on scalar optical fields just giving Doppler frequency shift. In this article, the unified Doppler frequency shift and Doppler phase shift in new vectorial Doppler effect have been demonstrated to allows for codetermination of the full information of motion vector (velocity magnitude and direction).

From a physical perspective, the direction-distinguishability of velocity vector or pseudovector (axial vector) based on this vectorial Doppler effect can be associated with the time-reversal symmetry (T-symmetry) breaking of Doppler signals from the classically entangled light fields. As is well known, under the transformation of time reversal, i.e. $T: t \rightarrow -t$, the velocity vector reverses its symbol (direction), i.e. $v \rightarrow -v$, being the odd symmetric. This implies that the T-symmetry of Doppler signals has to be broken in order to distinguish the direction of velocity vector. We have presented that the classical Doppler effects based on scalar optical fields cannot break the T-symmetry of one-dimensional Doppler intensity signals, thus being incapable of distinguishing the direction of velocity vector directly (see supplementary materials).

As for the spatially variant polarized light fields that feature a classical polarization-spatial entanglement[35-38], when a targeted particle moves within them and samples (reflects or scatters) out of the local polarization from the spatially variant polarized fields, it will transform the classical entanglement from the polarization-spatial entangled state to a new polarization-time entangled state (see supplementary materials). The resulting two-dimensionally polarization-time entangled DPSs are completely T-asymmetric, giving rise to the chirality inversion of DPS when reversing the motion direction of the particle as the result of the time reversal (see Figs. 1b and 1d). Such T-symmetry breaking of vectorial Doppler effect resembles the case of T-symmetry broken by magnetic fields that results in the symbol inversion of Lorentz force on a moving charged particle when having opposite motion directions.

In the proof-of-principle experiments, we just presented the detection of rotational motion of the targeted particle. It is worth mentioning that, naturally, this detection scenario can be extended to arbitrary motion states and even 3D motion tracking by using arbitrarily tailorable spatial polarization fields. This vectorial Doppler metrology approach enables the direct and robust detection of motion vector, because without an additional reference light, and thus exhibits high anti-interference to environment disturbance. This benefits from the exploitation of the spatially variant polarized light fields as an integration of spatial phase and polarization degrees of freedom of light (or classical polarization-spatial entanglement), in contrast to the simplex phase degree of freedom of scalar optical fields used in the previous scheme[6,15]. Our findings may offer many emerging applications in Doppler velocimetry, metrology, and monitoring for universal motion vectors in the natural world and human industry.

**Methods**

**Real-time monitoring.** A real-time monitoring for instantaneous position and velocity of a particle with variable motion can be achieved by directly counting the peaks in the Doppler intensity spectra and meanwhile estimating the Doppler phase shifts between two linearly polarized Doppler components. In general, for the spatially variant polarized light fields with a spatial period $\Lambda = \pi/k\gamma$ along the $x$ axis, one can track the instantaneous displacement of a moving particle relative to the initial position $x(t_0)$, as follows,

$$x(t_{m+1}) = \begin{cases} x(t_m) + \dfrac{\pi}{k\gamma}, & \Delta\varphi = -2\sigma\Delta\theta \\ x(t_m) - \dfrac{\pi}{k\gamma}, & \Delta\varphi = 2\sigma\Delta\theta \end{cases}, \quad (4)$$

where $m = 0, 1, 2\ldots$ denotes the order number of peaks found in the Doppler intensity signals. $t_m$ corresponds to the moment of the m$^{th}$ intensity peak, and $t_0$ the starting time of counting peaks. The resulting instantaneous velocity vector at the moment of $t_{m+1}$ can be given as

$$\vec{v}(t_{m+1}) = \dfrac{\vec{v}(t_{m+1}) - \vec{v}(t_m)}{\Delta t_{m+1}} = \begin{cases} \dfrac{\pi}{\Delta t_{m+1}\gamma}, & \Delta\varphi = -2\sigma\Delta\theta \\ -\dfrac{\pi}{\Delta t_{m+1}\gamma}, & \Delta\varphi = 2\sigma\Delta\theta \end{cases}, \quad (5)$$

where $\Delta t_{m+1} = t_{m+1} - t_m$ is the time interval between two adjacent intensity peaks.

As for the real-time monitoring of rotational movement of a particle using cylindrical VPFs in the experiment, the small rotational angle can be approximated as $d\phi = dx/r$, where $dx$ denotes the small displacement along the tangential direction, and the skew angle $\gamma = \ell/kr$. Thus from Eq. (4), the instantaneous angular position of the moving particle relative to its initial position $\phi(t_0)$ can be given as,

$$\phi(t_{m+1}) = \begin{cases} \phi(t_m) + \dfrac{\pi}{\ell}, & \Delta\varphi = -2\sigma\Delta\theta \\ \phi(t_m) - \dfrac{\pi}{\ell}, & \Delta\varphi = 2\sigma\Delta\theta \end{cases}, \quad (6)$$

as well as the resulting angular velocity,

$$\vec{\Omega}(t_{m+1}) = \frac{\phi(t_{m+1}) - \phi(t_m)}{\Delta t_{m+1}} = \begin{cases} \dfrac{\pi}{\Delta t_{m+1}\ell}, & \Delta\varphi = -2\sigma\Delta\theta \\ -\dfrac{\pi}{\Delta t_{m+1}\ell}, & \Delta\varphi = 2\sigma\Delta\theta \end{cases}. \quad (7)$$

**Generating cylindrical VPF and mimicking a rotating particle.** In the experiment, we took the means of a Sagnac interferometer configuration to generate the high-quality cylindrical VPFs by superposing two twisted light beams with opposite topological charge number ($\pm\ell$) and circular polarization $\sigma = \pm 1$ (see supplementary materials). Using this method, before superposing into the desired VPFs outside of the sagnac loop, these two light beams along the opposite directions in the loop propagate the same light paths, and thus can avoid the phase variation due to the ambient interference so as to generate the stable and high-quality cylindrical VPFs. The particle was mimicked by a lump of 78 adjacent micromirrors of a digital micromirror device (DMD) in the 'ON' state with a diameter of about 137 μm. The rotational movement of the particle was gotten by controlling the time interval to switch the next set of micromirrors to the 'ON' state (see supplementary materials). During the measurement, when illuminating the rotating particle, the pattern of the cylindrical VPF was controlled to match the rotational trajectory of the particle with the rotation diameter about 2.5 mm to obtain the distinguishable DPS.

**Acknowledgments:**

This work was supported by the National Natural Science Foundation of China (NSFC) (11774116, 11574001, 61761130082), the National Basic Research Program of China (973 Program) (2014CB340004), the Royal Society-Newton Advanced Fellowship, the National Program for Support of Top-notch Young Professionals, the Yangtze River Excellent Young Scholars Program, the Natural Science Foundation of Hubei Province of China (2018CFA048), the Key R&D Program of Guangdong Province (2018B030325002), and the Program for HUST Academic Frontier Youth Team (2016QYTD05).

**Author contributions:**

J.W. and L.F. developed the concept and conceived the experiments. Z.W. carried out the experiments and acquired the experimental data. L.F. performed the theoretical analyses. Z.W. and L.F. carried out the data analysis. L.F. contributed to writing the paper. J.W. finalized the paper. J.W. supervised the project.

**Supplementary Materials:**

Materials and Methods

Supplementary Text

Figures S1-S6

Supplementary Materials for

# A vectorial Doppler effect with spatially variant polarized light fields

Liang Fang[1+], Zhenyu Wan[1+], Andrew Forbes[2], Jian Wang[1*]

[1]Wuhan National Laboratory for Optoelectronics, School of Optical and Electronic Information, Huazhong University of Science and Technology, Wuhan 430074, Hubei, China.

[2]School of Physics, University of the Witwatersrand, Private Bag 3, Johannesburg 2050, South Africa

[+] These author contributed equally to this work.

[*] Correspondence to: jwang@hust.edu.cn

**This PDF file includes:**

Materials and Methods

Supplementary Text

Figures S1-S6

## Classical Doppler effect based on scalar optical fields

In the classical Doppler velocimetry based on scalar optical fields, two light beams as cross-reference are usually used for illumination on a moving object, as shown in Fig. S1a. For the convenience of analysis, we assume that the direction of observation is perpendicular to the motion direction. Firstly, we derive the field distribution of interference between two cross-reference light beams along the motion direction of a moving particle. Generally, the complex electric-field function of one light beam projecting in the motion direction can be written by,

$$E = A_0(r) \cdot \exp\left[i\left(-\omega t + \vec{k} \cdot \vec{x}\right)\right], \tag{S1}$$

where $A_0(r)$ is the complex amplitude of electric-field as a function of the spatial position, and $\omega$ is the angular frequency of light. $\vec{k}$ is the wave vector. For the conventional plane wave, the amplitude of the wave vector is the wavenumber defined as $k = 2\pi/\lambda$ with $\lambda$ being the wavelength, and its direction determines the propagation direction of light or Poynting vector. Here the particle moves along the $\vec{x}$ direction.

As for one of the cross-reference light beams, this projected electric-field in the region of interference can be further written by,

$$E_j = A_j(r) \cdot \exp\left[i\left(-\omega t + k \cdot x \cdot \cos\beta_j\right)\right], \tag{S2}$$

where $j = 1, 2$ represents the beam1 and beam2, respectively. $\beta_j$ denotes the angle between the wave vector $\vec{k}$ and the motion direction $\vec{x}$. For simplicity, here we set $\beta_1 = \pi/2 - \gamma$ for beam1, and $\beta_2 = \pi/2 + \gamma$ for beam2, where $\gamma$ is a small angle ($\gamma \approx \sin\gamma$) between the direction of incident light beams and the normal (observation) direction with respect to the trajectory of the moving particle. Thereby, the resulting interference fields between these two cross-reference light beams in Fig. S1b can be written by,

$$I = (E_1 + E_2) \cdot (E_1 + E_2)^* \approx 2|A_0|^2 \cdot \left[1 + \cos(2k \cdot x \cdot \gamma)\right], \tag{S3}$$

where the superscript '$*$' stands for complex conjugation. Note that here the complex amplitudes of two cross-reference light beams are approximated as $A_1 \approx A_2 \approx A_0$. From Eq. (S3), the period of interference fringes can be deduced as $\Lambda = \lambda/2\gamma$ indicated in Fig. S1b.

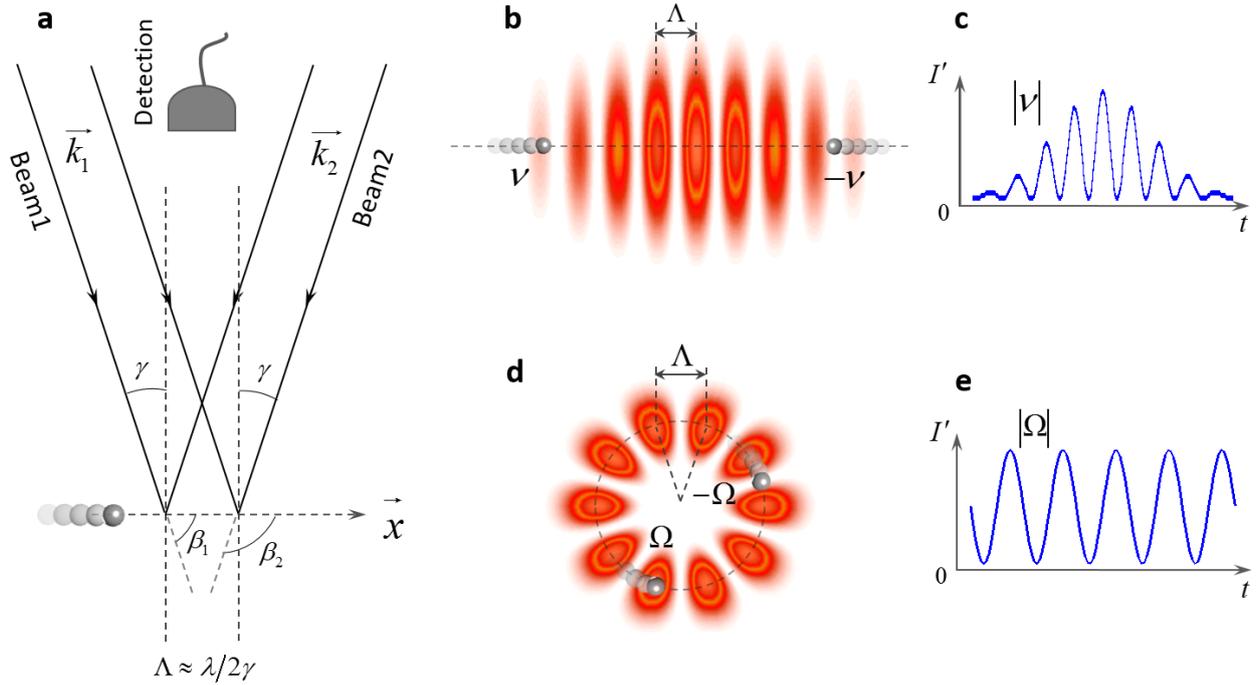

**Fig. S1** Classical Doppler effect based on scalar optical fields in the conventional Doppler velocimetry. **a** Two cross-reference light beams are used for detection of translational or rotational motion of a moving particle. **b** Two plane-phase beams produce straight interference fringes for interaction with translational motion of the particle. **c** Doppler signal of the scattered light by the moving particle with translational motion in interference fields corresponding to **b**. **d** Two twisted-phase beams produce rotational interference fringes for interaction with rotational motion of the particle. **e** Doppler signal of the scattered light by the moving particle with rotational motion in interference fields corresponding to **d**. Note that the motion direction of the moving particle cannot be distinguished based on the one-dimensional Doppler signals in **c** and **e** that are direction ambiguous because of their inherent time-reversal symmetry (T-symmetry) property.

When a particle moves within the interference fields at the velocity of $\vec{v}$ along the $\vec{x}$, it can scatter light into the detector. Because of normal detection with respect to the motion direction, the Doppler shift of the detected light is just induced by the interaction between the moving particle and the projecting fields expressed by Eq. (S1). In this case, the displacement $\vec{x}$ can be replaced by $\vec{x}=\vec{v}\cdot t$, so from Eq. (S1), the frequency of the scatted light becomes $\omega'=\omega-\vec{k}\cdot\vec{v}$, and thus the Doppler shift is $\Delta\omega=\omega'-\omega=-\vec{k}\cdot\vec{v}$. Note that this Doppler shift has a sign of plus or minus corresponding to

Doppler blue or red shift, respectively, which is determined by the relative direction of velocity $\vec{v}$ with respect to the wave vector $\vec{k}$.

Because of the ultra-high angular frequency $\omega$ of light, the Doppler shift is usually indirectly extracted by the interference with a reference wave. Here, the local frequency $\omega$ is removed by the beating of two cross-reference light beams, leaving the relatively low frequency Doppler shift. When considering the synchronous interaction of the moving particle with two projecting fields of the cross-reference light beams ($j=1$ and 2) given by Eq. (S2), substituting $x$ with $x=v\cdot t$, the Doppler signal of the scattered light from interference fields given by Eq. (S3) can be written by,

$$I'(t) \approx 2|A|^2 \left[1+\cos(2kv\gamma t)\right], \tag{S4}$$

where $A$ denotes the complex amplitude of the scattering light into the detector from each of two cross-reference light beams. Thereby, by fast Fourier transform (FFT) for this Doppler signal, the classical Doppler shift can be extracted as $\Delta f = k\cdot v\cdot \gamma/\pi$.

It should be stressed that the case of classical Doppler velocimetry used for detection of the above translation motion can also be extended to the rotational Doppler velocimetry when superposing two beams with orthogonally twisted components $\pm\ell$, where $\ell$ represents the topological charge number in Figs. S1d and S1e. In this case, the wave vector $\vec{k}$ should be modified by adding an azimuthal component, i.e. $\vec{k} \rightarrow \ell/r\cdot\vec{\phi}+k\cdot\vec{z}$ in cylindrical coordinate, giving a new wave vector with a skew angle $\gamma = \ell/kr$, with respect to the conventional wave vector, where $r$ is the radial position away from the axis. Note that the linear motion can be linked to the rotational motion by a relationship of $v=r\Omega$. According to the Doppler signal given by Eq. (S4), linear Doppler shift is evolved into the rotational Doppler shift, i.e. $\Delta f = k\cdot v\cdot \gamma/\pi = \ell\Omega/\pi$.

When taking the time-reversal transformation for the Doppler signal in Eq. (S4), the resulting Doppler signal becomes,

$$I'(-t)=I'(t) \approx 2|A|^2 \left[1+\cos(2kv\gamma t)\right]. \tag{S5}$$

Obviously, the classical Doppler effect based on scalar optical fields shows time-reversal symmetry (T-symmetry) of the Doppler intensity signal. This makes the directional ambiguity of detecting the moving object with either linear or rotational motion by the corresponding linear or rotational Doppler effect. Hence, the one-dimensional Doppler signals in Figs. S1c and S1e do not carry any direction information of the moving object. Thus, the classical Doppler effect based on scalar optical fields can be regarded as scalar Doppler effect.

**Vectorial Doppler effect based on vectorial polarization fields (VPFs)**

The classical Doppler effect by scalar optical fields discussed above is not associated with the polarization degree of freedom. In this section, we take consideration of the polarization into the system of Doppler velocimetry. By the same way as the analysis of scalar Doppler effect above, firstly we present how to produce the VPFs and then discuss how a moving object interacts with the VPFs, as illustrated in Fig. S2a. As for one of two cross-reference light beams with orthogonal circular state of polarization (SoP), the polarized electric-fields of light projecting in the motion direction of the moving object can be expressed using Jones vector as follows,

$$\vec{E}_j = A_j(r) \cdot \begin{bmatrix} 1 \\ \sigma_j i \end{bmatrix} \exp\left[i\left(k \cdot x \cdot \cos\beta_j + \chi_j\right)\right], \tag{S6}$$

where $j = 1$ and 2 denotes the beam1 and beam2, respectively, $A_j(r)$ is the real amplitude of electric-field, and $\chi_j$ is the initial phase of light beams. Note that here $\sigma = +1$ and $-1$ describes the left-handed and right-handed circular SoP, respectively, and other parameters are the same as those in Eq. (S2).

Along the motion direction of a moving object, these two cross-reference light beams can superpose into the VPFs in Fig. S2b as follows,

$$\vec{E} = \vec{E}_1 + \vec{E}_2 = \left\{ \begin{array}{l} A_1(r) \cdot \begin{bmatrix} 1 \\ \sigma_1 i \end{bmatrix} \exp\left[i(k \cdot x \cdot \cos\beta_1 - \alpha)\right] \\ + A_2(r) \cdot \begin{bmatrix} 1 \\ \sigma_2 i \end{bmatrix} \exp\left[i(k \cdot x \cdot \cos\beta_2 + \alpha)\right] \end{array} \right\} \exp\left[\frac{1}{2}i(\chi_2 + \chi_1)\right] \approx A_0(r) \cdot \begin{bmatrix} \cos(kx\gamma - \alpha) \\ -\sigma \cdot \sin(kx\gamma - \alpha) \end{bmatrix}$$

, (S7)

where the phase difference $\alpha = (\chi_2 - \chi_1)/2$ determines the initial polarization orientation of VPFs, and the spatial functions with respect to real amplitudes are also approximated as $A_1 \approx A_2 \approx A_0/2$, and $\sigma_1 = -\sigma_2 = \sigma$. Note that the phase term of $\exp[i(\chi_2 + \chi_1)/2]$ in Eq. (S7) is disregarded when deriving the final expression, because it does not affect the polarization distribution of VPFs. Even though the synthesized VPFs (superposed field) shown Fig. S2b may not stably propagate over a long distance in free space because of not satisfying Maxwell's equations, it does not affect the analysis of general VPFs to realize determination of motion vector as a conceptual illustration of vectorial Doppler velocimetry. For these VPFs, the period of spatial polarization variation is also $\Lambda = \pi/k\gamma = \lambda/2\gamma$. Similar to Fig. S1d, for the case of superposition of two twisted-phase light beams, $\gamma = \ell/kr$ and $x = r\phi$, where $\phi$ is azimuthal position in the cylindrical coordinate, the general VPFs expressed in Eq. (S7) are simplified to the cylindrical VPFs,

$$\vec{E} \approx A_0(r) \cdot \begin{bmatrix} \cos(\ell\phi - \alpha) \\ -\sigma \sin(\ell\phi - \alpha) \end{bmatrix}, \quad (S8)$$

where $\sigma = +1$ describes the cylindrical VPFs analogous to $HE_{\ell+1,1}$ vector mode, whereas $\sigma = -1$ denotes those analogous to $EH_{\ell-1,1}$ vector mode, as shown in Fig. S2d. Note that it features the spatial polarization orientation inversion between these two kinds of cylindrical VPFs ($\sigma = +1$ and 1).

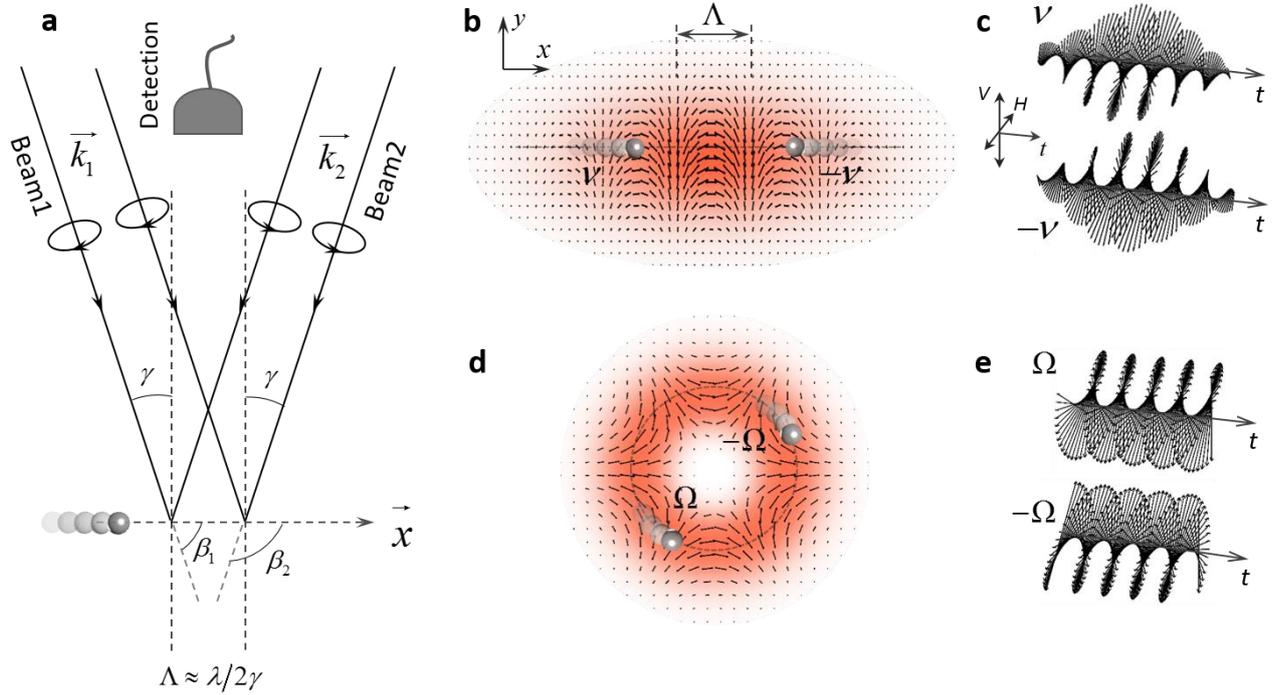

**Fig. S2** Vectorial Doppler effect based on vectorial polarization fields (VPFs). **a** Two cross-reference light beams with orthogonal circular state of polarization (SoP) are used for detection of velocity vector or pseudovector (magnitude, direction) of a moving particle. **b** Two plane-phase beams superpose into the VPFs along the x direction for interaction with translational motion vector of the particle. **c** Doppler polarization signals (DPSs) of the scattered light by the moving particle with translational motion vector (opposite directions) in VPFs corresponding to **b**. **d** Two twisted-phase beams superpose into the cylindrical VPFs for interaction with rotational motion vector of the particle. **e** DPS of the scattered light by the moving particle with rotational motion vector (opposite directions) in cylindrical VPFs corresponding to **d**. Note that the motion direction of the moving particle can be distinguished based on the two-dimensional DPSs in **c** and **e** that break the T-symmetry and lead to chirality inversion when reversing the velocity vector under the transformation of time reversal.

In accordance with section 1, also considering a particle moving at the velocity of $\vec{v}$ along the $\vec{x}$ within the general VPFs given by Eq. (S7), as shown in Fig. S2a, it will scatter the polarized light into the detector. The detected polarized light as two-dimensional Doppler polarization signal (DPS) can be written by,

$$\vec{E}(t) \approx A \cdot \begin{bmatrix} \cos(k\gamma vt - \alpha) \\ -\sigma \cdot \sin(k\gamma vt - \alpha) \end{bmatrix}, \quad (S9)$$

where $A$ is the real amplitude of the detected polarized light. Taking the time-reversal transformation for this DPS, the resulting DPS becomes,

$$\vec{E}(-t) \approx A \cdot \begin{bmatrix} \cos(k\gamma vt + \alpha) \\ \sigma \cdot \sin(k\gamma vt + \alpha) \end{bmatrix}. \quad (S10)$$

One can clearly see $\vec{E}(-t) \neq \vec{E}(t)$. The T-symmetry breaking of DPS between Eq. (S9) and Eq. (S10) is associated with the opposite chirality of DPS when reversing the motion direction of the particle within the VPFs under the time-reversal transformation, as shown in Figs. S2c and S2e. Apparently, the chiral DPS based on VPFs is a two-dimensional signal, carrying full information (magnitude, direction) of motion vector, belonging to the distinct feature of the vectorial Doppler effect. By contrast, the direction of the motion vector is not achievable by the one-dimensional Doppler intensity signal from the classical scalar Doppler effect, as shown in Figs. S1c and S1e.

# Classical entanglement transformation from polarization-spatial to polarization-time entangled states

Entanglement is studied almost exclusively in the context of quantum systems. This concept was initially introduced by Schrödinger[1] in his response to the famous gedanken experiment of Einstein, Podolsky, and Rosen (EPR)[2]. In recent years, the term entanglement has come to be used in a more general context[3-7]. A profound difference exists between two types of entanglement: (i) entanglement between spatially separated systems and (ii) entanglement between different degrees of freedom of a single system[4,5]. Only type (i) is identified with true entanglement in quantum physics that refers to nonlocal correlations. Conversely, type (ii) can be regarded as classical entanglement that appears in classical systems and cannot generate nonlocal correlations.

In a general paraxial beam as a single system, its polarization and spatial degrees of freedom can be coupled into four-dimensional vectors,

$$\vec{E} = E_{00}\vec{e}_R f_{10}(r) + E_{01}\vec{e}_R f_{01}(r) + E_{10}\vec{e}_L f_{10}(r) + E_{11}\vec{e}_L f_{01}(r), \quad (S11)$$

where $\vec{e}_R = 1/\sqrt{2}(\vec{e}_H - i\vec{e}_V)$ and $\vec{e}_L = 1/\sqrt{2}(\vec{e}_H + i\vec{e}_V)$ stand for the right-handed and left-handed circular SoP of light, respectively, $f_{10}(r)$ and $f_{01}(r)$ represent the spatial functions. Thereby, the vectors in Eq. (S11) can be represented as the ket $|\psi\rangle \doteq [E_{00}, E_{01}, E_{10}, E_{11}]^T$ with T denoting matrix transposition. These four-dimensional vectors can be regarded as existing in a four-dimensional two-qubit Hilbert space: $H = H_{pol} \otimes H_{spa}$, where $H_{pol} = \text{span}\{\vec{e}_R, \vec{e}_L\}$ denotes the polarization-qubit space, and $H_{spa} = \text{span}\{f_{10}(r), f_{01}(r)\}$ the spatial-qubit space. Following Ref. (5), we identify the standard basis for the polarization qubit with right- and left-handed circular SoP and the standard basis for the spatial qubit as follows,

$$|R\rangle \doteq \vec{e}_R, \quad |L\rangle \doteq \vec{e}_L, \quad (S12)$$

$$|f_{10}\rangle \doteq f_{10}(r), \quad |f_{01}\rangle \doteq f_{01}(r). \quad (S13)$$

With these notations, the vectors in Eq. (S11) in this two-qubit Hilbert space H can be rewritten by,

$$|\psi(r)\rangle = E_{00}|R\rangle|f_{10}\rangle + E_{01}|R\rangle|f_{01}\rangle + E_{10}|L\rangle|f_{10}\rangle + E_{11}|L\rangle|f_{01}\rangle, \quad (S14)$$

As for the system of VPFs given by Eq. (S7), the simplified expression just in terms of polarization and spatial degrees of freedoms can be expressed as,

$$\vec{E} \approx \frac{A_0(r)}{2} \cdot \left\{ \begin{bmatrix} 1 \\ \sigma i \end{bmatrix} \exp[i(kx\gamma - \alpha)] + \begin{bmatrix} 1 \\ -\sigma i \end{bmatrix} \exp[-i(kx\gamma - \alpha)] \right\}. \tag{S15}$$

In this case, spatial functions $f_{10}(r) = A_0(r) \cdot \exp[i(kx\gamma - \alpha)]$ and $f_{01}(r) = A_0(r) \cdot \exp[-i(kx\gamma - \alpha)]$, and $E_{00} = E_{11} = 0$, $E_{01} = E_{10} = 1/\sqrt{2}$, $\sigma = +1$, VPFs in Eq. (S7) can be described with the state

$$|\psi_1(r)\rangle = \frac{1}{\sqrt{2}}(|L\rangle|f_{10}\rangle + |R\rangle|f_{01}\rangle), \tag{S16}$$

or $E_{00} = E_{11} = 1/\sqrt{2}$, $E_{01} = E_{10} = 0$ and $\sigma = -1$ giving the state

$$|\psi_2(r)\rangle = \frac{1}{\sqrt{2}}(|L\rangle|f_{01}\rangle + |R\rangle|f_{10}\rangle), \tag{S17}$$

The states in Eqs. (S16) and (S17) correspond to two kinds of spatial polarization orientation of VPFs, clearly, both showing that the polarization and spatial degrees of freedom for VPFs are nonseparable in terms of classical entanglement. Because these fields cannot be represented as a simple product of one polarization vector and one spatial function. In accordance with section 2, when $\gamma = \ell/kr$ and $x = r\phi$, the VPFs expressed in Eqs. (S16) and (S17) reduce to the familiar case of cylindrical VPFs as classical polarization-spatial entanglement[5,6].

In the system of vectorial Doppler velocimetry, when scattered by a moving particle, the DPSs from the classical polarization-spatial entangled VPFs can be approximately given by

$$\vec{E}(t) \approx \frac{A}{2} \cdot \left\{ \begin{bmatrix} 1 \\ \sigma i \end{bmatrix} \exp[i(k\gamma vt - \alpha)] + \begin{bmatrix} 1 \\ -\sigma i \end{bmatrix} \exp[-i(k\gamma vt - \alpha)] \right\}, \tag{S18}$$

where the SoP that varies as a function of time cannot be separated as a simple product of one polarization vector and one time-dependent function. Obviously, the polarization-spatial entangled states in Eqs. (S16) and (S17) are transferred into polarization-time entangled states by replacing the spatial functions $f_{10}(r)$ and $f_{01}(r)$ with the corresponding time-dependent functions $h_{10}(t) = A \cdot \exp[i(k\gamma vt - \alpha)]$ and $h_{01}(t) = A \cdot \exp[-i(k\gamma vt - \alpha)]$. Hence, the polarization-spatial entanglement in Eq. (S16) becomes,

$$|\psi_1(t)\rangle = \frac{1}{\sqrt{2}}(|L\rangle|h_{10}\rangle + |R\rangle|h_{01}\rangle), \tag{S19}$$

and the entanglement in Eq. (S17) corresponds to

$$|\psi_2(t)\rangle = \frac{1}{\sqrt{2}}(|L\rangle|h_{01}\rangle + |R\rangle|h_{10}\rangle), \tag{S20}$$

In quantum mechanics, the time-reversal operator $\hat{T}$ for wave functions is described by $\hat{T}: \psi(t) \to \psi^*(-t)$ [8]. Under this operation, the states in Eqs. (S19) and (S20) become reciprocally time-reversal entangled states, namely,

$$\hat{T}|\psi_1(t)\rangle = |\psi_2(t)\rangle, \quad \hat{T}|\psi_2(t)\rangle = |\psi_1(t)\rangle. \tag{21}$$

Note that the phase term of $\exp(i\alpha)$ here is not considered, since the wave function in different phase factors also belongs to the state of system determined by the same Schrödinger equation according to the linearity relation.

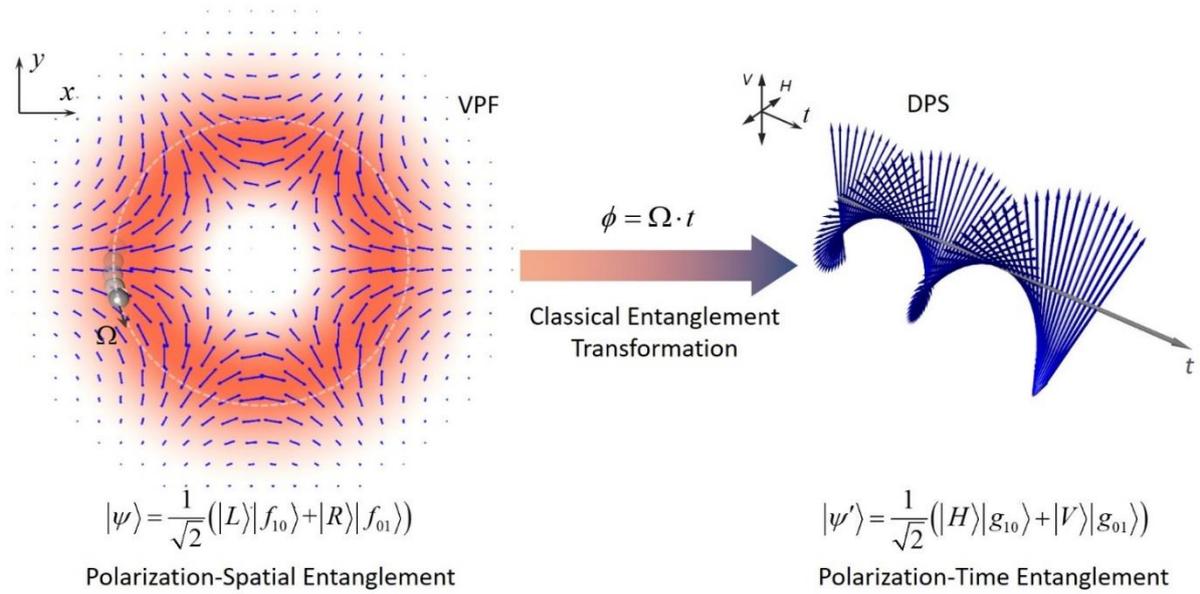

$$|\psi\rangle = \frac{1}{\sqrt{2}}(|L\rangle|f_{10}\rangle + |R\rangle|f_{01}\rangle)$$
Polarization-Spatial Entanglement

$$|\psi'\rangle = \frac{1}{\sqrt{2}}(|H\rangle|g_{10}\rangle + |V\rangle|g_{01}\rangle)$$
Polarization-Time Entanglement

**Fig. S3** Classical entanglement transformation (polarization-spatial to polarization-time) in the system of vectorial Doppler velocimetry based on VPFs.

From another perspective on the basis of horizontal (H) and vertical (V) polarizations, the DPS in Eq. (S18) can be also identified with

$$\vec{E}(t) \approx \frac{A}{2} \cdot \left[\vec{e}_H \cos(k\gamma vt - \alpha) \mp \vec{e}_V \sin(k\gamma vt - \alpha)\right]. \tag{S22}$$

Similarly, the new basis for the polarization and time qubit can be given by

$$|H\rangle \doteq \vec{e}_H, \quad |V\rangle \doteq \vec{e}_V, \tag{S23}$$

$$|g_{10}\rangle \doteq g_{10}(t) = \frac{1}{\sqrt{2}} A\cos(k\gamma vt - \alpha), \quad |g_{01}\rangle \doteq g_{01}(t) = -\frac{1}{\sqrt{2}} A\sin(k\gamma vt - \alpha). \tag{S24}$$

Despite being real functions, the time functions are also expressed by the kets as an analogical wave function. Consequently, the polarization-time entangled DPS can be rewritten by

$$|\psi'(t)\rangle = \frac{1}{\sqrt{2}}(|H\rangle|g_{10}\rangle \pm |V\rangle|g_{01}\rangle), \tag{S25}$$

and the form of time-reversal transformation,

$$\hat{T}|\psi'(t)\rangle = \frac{1}{\sqrt{2}}(|H\rangle|g_{10}\rangle \mp |V\rangle|g_{01}\rangle), \tag{S26}$$

From Eq. (S21) or Eq. (S26), and Fig. S3, obviously the T-symmetry of DPS is completely broken for one specific state of classical polarization-time entanglement in Eqs. (S19) and (S20). From perspective in quantum mechanics, as for the system of Doppler velocimetry based on VPFs, in a sense, the DPS with T-symmetry breaking can be explained by the introduction of spin (circular SoP) of light for the Hamiltonian system that necessarily changes the time-reversal state function.

**Experimental setup and the generation of VPFs for the detection of motion vector**

We built an experimental setup for the detection of motion vector using vectorial Doppler effect based on VPFs, as illustrated in Fig. S4a. The generation of VPFs was incorporated into the experimental setup. A He-Ne laser beam at 632.8 nm was shaped into a perfect Gaussian beam by propagating it through a piece of single-mode fiber (SMF). A group of polarizer (Pol.1) and half-wave plate (HWP1) was used to adjust the SoP of light beam to be $45^0$. When passing through a beam splitter (BS1) and a polarization beam splitter (PBS), the $45^0$ polarized light beam was spilt into x and y-polarized components in two paths. The light beam in each path was modulated into a twisted-phase beam by a spatial light modulator (SLM) in a Sagnac interferometer configuration. The HWP2 in the Sagnac interferometer configuration was used to rotate the linear SoP of light beams in both paths to its orthogonal state so that each light beam was modulated and then combined into a superposed beam through the PBS. Considering the reflection with odd and even times respectively for the two generated twisted-phase beams by the same SLM, the superposed beam contained two twisted-phase components with opposite topological charge number ($\pm\ell$) in x/y linear SoP. After the BS and passing through a quarter-wave plate (QWP1), the superposed beam becomes cylindrical VPFs.

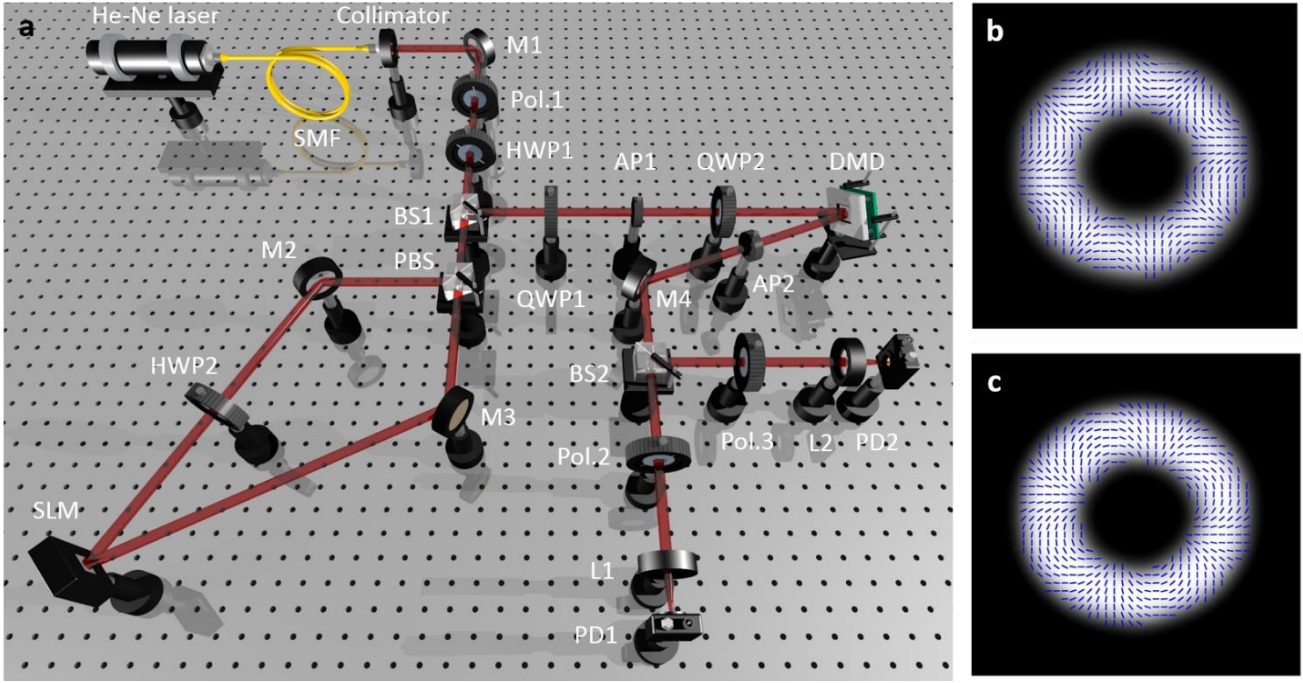

**Fig. S4** Experimental setup and the generation of VPFs for the detection of motion vector using vectorial Doppler effect. **a** Experimental setup. SMF: single-mode fiber; M: mirror; Pol.: polarizer; HWP: half-wave plate; BS: beam splitter; PBS: polarization beam splitter; SLM: spatial light modulator; QWP: quarter-wave plate; AP: aperture; DMD: digital micromirror device; L: lens; PD: photodetector. **b**, **c** Measured intensity and polarization distributions of the high-quality $HE_{41}$-like cylindrical VPFs in **b** and $EH_{21}$-like cylindrical VPFs in **c** generated in the experiment.

The Sagnac interferometer configuration sharing the same light path benefits the high-quality generation of cylindrical VPFs. In the experiment, cylindrical VPFs analogous to $HE_{41}$ vector mode in Fig. S4b and analogous to $EH_{21}$ vector mode in Fig. S4c were generated by imprinting with a computer generated phase profile ($|\ell|=3$) to SLM and rotating QWP1 to control circular SoP. The generated cylindrical VPF was illuminated on a moving particle that is mimicked through a digital micromirror device (DMD) by setting micromirrors in specific time-varying positions to the state of 'ON'. We turned a lump of 78 adjacent micromirrors to the 'ON' state to mimic a microcircle with a diameter of about 137 μm. The diameter of the cylindrical VPF was controlled to be about 2.5 mm to match the rotational radius of the moving particle. Otherwise the moving particle could interact with imprecise local SoP, making the DPS fuzzy and undistinguishable. In the experiment, the size ratio of the particle to the period of spatial SoP variation in VPF is about 0.1. The QWP2 is used for the compensation of the polarization-dependent dissipation when reflecting the local SoP by the mimicked particle consisting of an array of micromirrors. The apertures (AP1 and AP2) could block the light in undesired diffraction orders generated by SLM and DMD.

**Additional measured results and discussions**

In the main text, we presented the experimental results for detecting angular velocities under opposite directions using the cylindrical VPF analogous to $HE_{41}$ vector mode. As is well known, this cylindrical VPF consists of light components with the same twisted-phase basis as the $EH_{21}$ vector mode. The cylindrical VPF analogous to $EH_{21}$ vector mode can be generated simply by exchanging the circular SoP between two twisted-phase components used for synthesizing $HE_{41}$-like VPF. In the experiment, we also employed $EH_{21}$-like VPF to detect the same angular velocities, as shown in Fig. S5. Compared to the measured results using $HE_{41}$-like VPF in Figs. 2d and 2e, the measured two Doppler intensity signals using $EH_{21}$-like VPF in Figs. S5a and S5d just reversed the relative phase difference that was clearly shown in the relative phase spectra in Figs. S5c and S5f. The full information (magnitude, direction) detection of motion vector is also achievable using $EH_{21}$-like VPF.

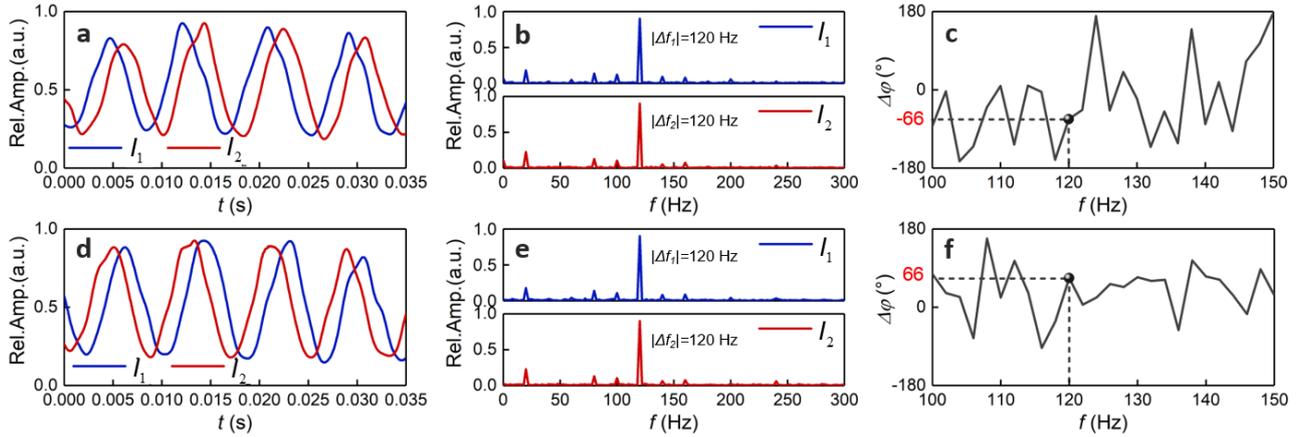

**Fig. S5** Measured results for vectorial Doppler effect based on cylindrical VPFs analogous to $EH_{21}$ vector mode. **a**,**d** Measured Doppler intensity signals by two photodetectors after filtering the DPS through two polarizers. **b**, **c**, **e**, **f** Doppler frequency spectra acquired by fast Fourier transform (FFT) of the recorded Doppler intensity signals in **a** and **d**. **b**, **e** Amplitude spectra. **c**, **f** Relative phase spectra showing relative phase difference between two path frequency signals. **a-c** $\vec{\Omega}=40\pi$ rad/s. **d-f** $\vec{\Omega}=-40\pi$ rad/s.

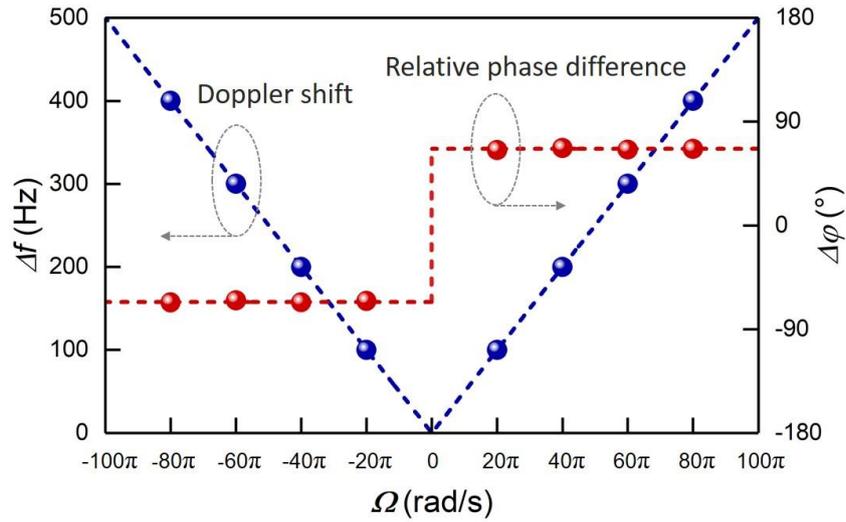

**Fig. S6** Measured results (Doppler frequency shift, relative phase difference) for vectorial Doppler effect based on cylindrically VPFs analogous to $EH_{61}$ vector mode under different angular velocities. The dots represent measured results and the dashed lines denote theories.

Moreover, we also experimentally demonstrated the variance of the Doppler shift and the invariance of the relative phase difference under different angular velocities of the rotating particle using the cylindrical VPF analogous to $EH_{61}$ vector mode, as shown in Fig. S6. It clearly shows that the Doppler shift is proportional to the magnitude of motion velocity, while the relative phase difference depends on the direction of motion vector.